\documentclass[12pt]{iopart}

\usepackage{iopams}
\usepackage{graphicx}
\expandafter\let\csname equation*\endcsname\relax
\expandafter\let\csname endequation*\endcsname\relax
\usepackage{amsmath}
\usepackage{bm}
\usepackage{hyperref}
\usepackage{graphics}
\usepackage{slashed}
\usepackage{amssymb}
\usepackage{url}
\usepackage[dvipsnames,svgnames,table]{xcolor}
\usepackage[T1]{fontenc}

\makeatletter
\newcommand{\tpitchfork}{%
  \vbox{
    \baselineskip\z@skip
    \lineskip-.52ex
    \lineskiplimit\maxdimen
    \m@th
    \ialign{##\crcr\hidewidth\smash{$-$}\hidewidth\crcr$\pitchfork$\crcr}
  }%
}
\makeatother

\usepackage[normalem]{ulem}
\usepackage{color}

\begin{document}

\title[Covariant TOV equations in EMSG]{Covariant Tolman-Oppenheimer-Volkoff equations in Energy-Momentum Squared Gravity}

\author{Eduardo Bittencourt${}^{1,2}$\footnote{Corresponding author} and Mariam Campbell${}^{3}$ and Peter K. S. Dunsby${}^{2,4,5}$ and Sergio E. Jor\'as${}^{6}$}
\address{${}^{1}$Institute of Physics and Chemistry, Federal University of Itajub\'a, Av.\ BPS 1303, Itajub\'a/MG, Brazil}
\address{${}^{2}$Department of Mathematics and Applied Mathematics, University of Cape Town, Rondebosch 7700, Cape Town, South Africa}
\address{${}^{3}$Astrophysics Research Centre, School of Agriculture and Science, University of KwaZulu-Natal, Private Bag X54001, Durban 4000, South Africa}
\address{${}^{4}$Center for Space Research, North-West University, Potchefstroom 2520, South Africa}
\address{${}^{5}$South African Astronomical Observatory, Observatory 7925, Cape Town, South Africa}
\address{${}^{6}$Instituto de F\'\i sica, Universidade Federal do Rio de Janeiro,\\
 CEP 21941-972 Rio de Janeiro, RJ, Brazil}
\ead{bittencourt@unifei.edu.br, CampbellM@ukzn.ac.za, peter.dunsby@uct.ac.za, joras@if.ufrj.br}

\date{\today}

\begin{abstract}

We study static, spherically symmetric stellar configurations in an extended class of Energy--Momentum Squared Gravity using the covariant \(1+1+2\) semi-tetrad formalism. For perfect physical fluids, we show that the nonlinear matter corrections can be reinterpreted as an effective perfect fluid, so that the stellar equilibrium equations retain the standard Tolman--Oppenheimer--Volkoff form when written in terms of effective variables. The resulting covariant structure equations are formulated in both metric and dimensionless variables and, whenever an effective closure relation exists, reduce to an autonomous planar dynamical system. This provides a global qualitative description of the stellar phase space in terms of finite and asymptotic critical points. 

Specializing to linear physical equations of state, we recover the general relativistic benchmark and identify sectors that are exactly, asymptotically, or piecewise equivalent to general relativity, as well as sectors---particularly dust configurations---for which the planar reduction breaks down and the full three-dimensional covariant flow must be considered. We further recover the standard metric Tolman--Oppenheimer--Volkoff equation in terms of effective variables and show that, although the exterior spacetime remains Schwarzschild, the natural matching condition at the stellar surface is \(p_{\rm eff}(R)=0\), which need not coincide with \(p(R)=0\) for self-bound matter.

\end{abstract}


\maketitle

\section{Introduction}
\label{sec:intro}

General Relativity (GR) has successfully passed a broad range of observational tests, from solar-system experiments to binary pulsars and gravitational-wave measurements. Nevertheless, several outstanding problems---including the nature of dark matter and dark energy, the resolution of cosmological singularities, and the possibility of deviations from GR in the strong-field regime---continue to motivate the study of alternative theories of gravity. Compact objects such as neutron stars and quark stars provide natural laboratories for probing strong gravitational fields and ultra-dense matter, making them particularly valuable for testing modifications of GR.

Among the many extensions of GR, a well-motivated class involves nonlinear couplings between gravity and matter scalars constructed from the energy-momentum tensor. In particular, \emph{Energy-Momentum Squared Gravity} (EMSG) is obtained by supplementing the Einstein--Hilbert action with a function of the scalar \(\mathcal{T}\equiv T_{\mu\nu}T^{\mu\nu}\)~\cite{Roshan:2016,Katirci:2014}. In EMSG, vacuum solutions coincide with those of GR, while matter sources acquire nonlinear self-interactions that become important at high energy densities and pressures. These corrections can generate nonsingular cosmological bounces, modify early-universe dynamics, and support viable cosmological histories~\cite{Board:2017,Akarsu:2018a}. The theory has also been constrained using compact stars and binary pulsars~\cite{Akarsu:2018b,Nazari:2022,Cipriano:2024}.

On cosmological scales, both the background evolution and the linear perturbations of EMSG can be reformulated in terms of effective fluid variables, with the nonlinear matter corrections absorbed into GR-like Friedmann and perturbation equations~\cite{Dunsby:2025}. In this framework, the effective equation-of-state parameter \(\bar{w}\) and effective sound speed \(\bar{c}_s^2\) may differ significantly from their GR counterparts at high densities. This effective-fluid interpretation provides a particularly useful way of comparing cosmological predictions of EMSG with observational constraints.

Stellar configurations in EMSG have so far been studied mainly within the standard metric approach, where modified Tolman--Oppenheimer--Volkoff (TOV) equations are derived and solved numerically for polytropic or realistic equations of state~\cite{Tolman:1939,Oppenheimer:1939,Akarsu:2018b,Nari:2018,Alam:2024,Banerjee:2025,Singh:2021,Pretel:2023,Nasir:2023,Sharif:2022}. These analyses typically retain the usual static, spherically symmetric metric ansatz and investigate how the modified hydrostatic equilibrium equations affect neutron-star masses, radii, and stability properties.

In parallel, covariant approaches based on semi-tetrad methods have provided an alternative formulation of relativistic stellar structure. The \(1+1+2\) covariant formalism~\cite{Clarkson:2003,Carloni:2018} offers a description of static, spherically symmetric spacetimes by recasting the TOV equations as a closed system of first-order propagation equations for a set of covariant scalar variables. This framework has been successfully applied to both single-fluid and two-fluid stellar models in GR~\cite{Naidu:2021}, and related covariant formulations have also been developed for static solutions in modified gravity theories such as \(f(R)\) gravity~\cite{Nzioki:2010,Campbell:2025}.

The purpose of the present work is to combine these developments by constructing static, spherically symmetric stellar models in EMSG within the \(1+1+2\) covariant formalism. In particular, we investigate whether the effective-fluid picture that emerges naturally in the cosmological sector of EMSG also extends consistently to relativistic stellar structure. We show that, for perfect physical fluids, the nonlinear EMSG corrections can indeed be recast as an effective perfect fluid, allowing the covariant stellar equations to retain the same structure as their GR counterparts when written in terms of effective energy density and pressure. This establishes a unified framework in which both cosmological and stellar tests of EMSG can be interpreted using the same effective-fluid description.

The paper is organized as follows. In Sec.~\ref{sec:theory_EMSG} we review the field equations of EMSG and the effective-fluid description for perfect fluids, considering the cases \(L_m=p\) and \(L_m=-\rho\). In Sec.~\ref{sec:112_formalism_TOV_EMSG} we summarize the \(1+1+2\) covariant formalism for static, spherically symmetric spacetimes and derive the corresponding covariant TOV equations in EMSG, together with a dimensionless formulation suitable for phase-space analysis. In Sec.~\ref{sec:qualitative_Kp_effective} we develop a qualitative dynamical-systems analysis of the reduced stellar equations, identifying nullclines, invariant sets, and equilibrium points. In Sec.~\ref{sec:phase_portraits_linearEOS} we specialize to linear physical equations of state, recovering the GR benchmark and analyzing the EMSG sectors for \(L_m=p\) and \(L_m=-\rho\), including radiation and dust configurations. Finally, in Sec.~\ref{sec:recovering_results} we recover the standard metric representation of the stellar equations, rewrite the system in terms of the physical variables, and discuss matching conditions and the definition of the stellar radius.

Throughout this work, we adopt geometric units with \(c=1\) and \(8\pi G=1\), together with metric signature \((-++\,+)\).

\section{Energy-Momentum Squared Gravity and effective fluid}
\label{sec:theory_EMSG}

\subsection{Action and field equations}

We consider the class of EMSG models defined by the action
\begin{equation}
S = \frac{1}{2}\int d^4x\,\sqrt{-g}\,F(R,\mathcal{T}) + \int d^4x\,\sqrt{-g}\,L_m,
\label{eq:action_general}
\end{equation}
with
\begin{equation}
F(R,T) = R + \eta\,\mathcal{T}^{n},\qquad \mathcal{T}\equiv T_{\mu\nu}T^{\mu\nu},
\label{eq:F_R_T_definition}
\end{equation}
where \(\eta\) is a coupling constant, \(n\in\mathbb{R}\), and \(L_m\) is the matter Lagrangian density. The energy-momentum tensor is defined by
\begin{equation}
T_{\mu\nu} = -\frac{2}{\sqrt{-g}}\,\frac{\delta(\sqrt{-g}\,L_m)}{\delta g^{\mu\nu}}.
\end{equation}
Varying the action with respect to the metric, yields~\cite{Roshan:2016,Dunsby:2025}
\begin{equation}
G_{\mu\nu} = T_{\mu\nu} + \eta\,\mathcal{T}^{n-1}\left[\frac{1}{2}\mathcal{T} g_{\mu\nu} - n\,\Theta_{\mu\nu} \right],
\label{eq:field_eq_EMSG}
\end{equation}
with
\begin{equation}
\Theta_{\mu\nu} \equiv -2L_m\left(T_{\mu\nu} - \frac{1}{2}T\, g_{\mu\nu}\right) - T\, T_{\mu\nu} + 2T_{\mu}{}^{\alpha}T_{\alpha\nu} - 4T^{\alpha\beta}\frac{\partial^2 L_m}{\partial g^{\mu\nu}\partial g^{\alpha\beta}},
\end{equation}
where $T\equiv g_{\mu\nu}T^{\mu\nu}$ is the trace of the energy-momentum tensor. In what follows, we shall adopt two approaches: the widely used choice \(L_m=p\) for perfect fluids, so that the last term above is \textit{neglected}, preventing possible ill-defined self-interacting terms~\cite{Harko:2011,Dunsby:2025}; and the less common case \(L_m=-\rho\), which provides a well-behaved contribution from the last term above. As discussed in Ref.\ \cite{Akarsu:2024}, both approaches are valid at the level of the field equations, although only the latter is safe for all kinds of matter at the Lagrangian level, particularly the case of dust.

Since the action is diffeomorphism-invariant, the total energy-momentum tensor on the right-hand side of Eq.~\eqref{eq:field_eq_EMSG} is covariantly conserved. However, the physical tensor \(T_{\mu\nu}\) is not conserved in general; instead, an effective interaction term mediates the exchange between the physical fluid and the EMSG corrections. This feature plays an important role in both cosmology and stellar structure, leading to different results from GR mainly at the high-energy regime.

\subsection{Perfect fluid and effective variables}

We assume that the matter content is a perfect fluid with normalized four-velocity \(u^\mu\), energy density \(\rho\), and pressure \(p\), whose energy-momentum tensor is given by
\begin{equation}
T_{\mu\nu} = \rho\,u_\mu u_\nu + p\,h_{\mu\nu},
\label{eq:Tmunu_perfect}
\end{equation}
where $u_\mu u^\mu=-1$ and $h_{\mu\nu}=g_{\mu\nu}+u_\mu u_\nu$ is the projector onto the 3-space orthogonal to $u_\mu$. In the fluid rest frame, one finds that
\begin{equation}
\label{eq:trace_squared}
\mathcal{T} = \rho^2 + 3p^2.
\end{equation}
From these quantities, we can define the \textit{physical} equation-of-state parameter $w\equiv p/\rho$ and the \textit{physical} sound speed $c_s^2\equiv \partial p/\partial\rho$, which are not necessarily constant.

Following the cosmological analysis of Ref.~\cite{Dunsby:2025}, it is convenient to define an \emph{effective} energy-momentum tensor \(T^{(\rm eff)}_{\mu\nu}\) such that the field equations take the GR-like form
\begin{equation}
G_{\mu\nu} = T^{(\rm eff)}_{\mu\nu}.
\label{eq:GR_like_EMSG}
\end{equation}
For a perfect fluid, the combination of \(T_{\mu\nu}\) and \(\Theta_{\mu\nu}\) in Eq.~\eqref{eq:field_eq_EMSG} can be written as a linear combination of \(u_\mu u_\nu\) and \(h_{\mu\nu}\). Consequently, \(T^{(\rm eff)}_{\mu\nu}\) retains the perfect-fluid form
\begin{equation}
T^{(\rm eff)}_{\mu\nu} = \rho_{\rm eff}u_\mu u_\nu + p_{\rm eff}\,h_{\mu\nu},
\label{eq:Teff_perfect}
\end{equation}
with effective energy density and pressure given by algebraic functions of \(\rho\) and \(p\),
\begin{align}
\rho_{\rm eff} &= \rho + \rho_{\rm emsg}(\rho,p;\eta,n),\\
p_{\rm eff}   &= p   + p_{\rm emsg}(\rho,p;\eta,n).
\end{align}
Because of the Bianchi identities, the effective tensor is conserved,
\begin{equation}
\nabla_\mu T^{\mu\nu}_{\rm (eff)} = 0,
\label{eq:Teff_conserved}
\end{equation}
so that it behaves as a genuine fluid source at the gravitational level compared to GR. The physical tensor \(T_{\mu\nu}\), on the other hand, satisfies a modified conservation law with a source term depending on \(\eta\), \(n\), \(\rho\), and \(p\), which gives a completely different interpretation of the components of the physical fluid when gravity is absent. In homogeneous cosmology, this leads to modified continuity equations and peculiar features for matter perturbations~\cite{Dunsby:2025}. In static stars, it will affect the relation between physical and effective pressure gradients.

For an EMSG model defined by Eq.\ \eqref{eq:F_R_T_definition}, the fluid for each Lagrangian case is described in the following way: When we consider the case $\mathcal{L}_m=p$, we explicitly obtain
\begin{eqnarray}
\rho_{\rm emsg} &=& \eta\,\mathcal{T}^{n-1}\left[\left(n-\frac{1}{2}\right)\mathcal{T}+4n\rho p\right],\label{eq:rho_eff_general_Lag_p}\\
p_{\rm emsg} &=& \frac{\eta}{2}\,\mathcal{T}^{n}.\label{eq:p_eff_general_Lag_p}
\end{eqnarray}
For a barotropic EoS for the physical fluid $p=p(\rho)$, one obtains closed-form expressions for the effective EoS parameter $w_{\rm eff}\equiv p_{\rm eff}/\rho_{\rm eff}$ and the effective adiabatic sound speed $[c_s^2]_{\rm eff}\equiv (dp_{\rm eff}/d\rho)(d\rho_{\rm eff}/d\rho)^{-1}$, as follows
\begin{align}
\label{eq:bar_w_Lp}
    w_{\rm eff} & = \frac{w + \frac{\eta}{2}(1 + 3w^2)^n\rho^{2n-1}}{1 + \eta A(n,w)\rho^{2n-1}},\\[1ex]
\label{eq:bar_cs2L_p}
    [c_s^2]_{\rm eff}  & = \frac{c_s^2 + \frac{1}{2}n\,\eta\,\mathcal{T}^{n-1}\, \mathcal{T}_{,\rho}}{1 + \eta\,\mathcal{T}^{n-1}\mathcal{T}_{,\rho} \left\{(n-1)\mathcal{T}^{-1}\left[\left(n-\frac{1}{2}\right)\mathcal{T}+4n\rho p\right] + \left(n-\frac{1}{2}\right) + \frac{4n(p + \rho c_s^2)}{\mathcal{T}_{\rho}}\right\}}.
\end{align}
where $A(n,w)=(1 + 3w^2)^{n-1}[(2n-1)(1 + 3w^2) + 8nw]/2$. We denote $d\mathcal{T}/d\rho\equiv\mathcal{T}_{,\rho}=2(\rho+3pc_s^2)$ and $\mathcal{T}_{,\rho\rho}=2[1+3c_s^4+3p(c_s^2)_{,\rho}]$ for brevity. For radiation ($p=\rho/3$) and $n=1$, one obtains $w_{\rm eff}=[c_s^2]_{\rm eff}=1/3$, which means that the physical and effective fluids possess the same parameters, although the physical and effective variables behave distinctly. A similar behavior occurs if $n=1/2$, in which case the EoS is linear, with $w_{\rm eff}=[c_s^2]_{\rm eff}=\mbox{const.}$, but $w_{\rm eff}>w$ for small $\eta>0$ or $w_{\rm eff}<w$ for small $\eta<0$. Note that branch degeneracies may appear near denominator zeros, as we shall see later on.

For $\mathcal{L}_m=-\rho$, we get
\begin{eqnarray}
\rho_{\rm emsg} &=& -\frac{\eta}{2}\,(\rho^2+3p^2)^{n},\label{eq:rho_eff_general_Lag_rho}\\
p_{\rm emsg}  &=& \frac{\eta}{2}\,(\rho^2+3p^2)^{n-1}\left[\rho^2+3p^2-2n(\rho+p)(\rho+3\,p\, c_s^2)\right]\label{eq:p_eff_general_Lag_rho},
\end{eqnarray}
with effective EoS parameter and effective adiabatic sound speed given now by
\begin{equation}
w_{\rm eff} = \frac{w+\eta B(n,w)\rho^{2n-1}}{ 1-\dfrac{\eta}{2}(1+3w^2)^n\rho^{2n-1}},
\label{eq:weff_Lmrho}
\end{equation}
and
\begin{equation}
[c_s^2]_{\rm eff} = \frac{ c_s^2+\dfrac{\eta}{2}\dfrac{d}{d\rho}
\left(\mathcal{T}^{\,n-1}{\cal Q}\right)
}{
1-\dfrac{\eta}{2}n\mathcal{T}^{\,n-1}\mathcal{T}_{,\rho}
},
\label{eq:ceff_Lmrho_compact}
\end{equation}
where $B(n,w) = \frac12(1+3w^2)^n[1-2n(1+w)]$, ${\cal Q}= \mathcal{T}-n(\rho+p)\mathcal{T}_{,\rho}$, and $\mathcal{T}_{,\rho}=2(\rho+3pc_s^2)$. Equivalently, if the derivative in Eq.~\eqref{eq:ceff_Lmrho_compact} is expanded along the physical barotropic EoS, one obtains
\begin{equation}
\frac{d}{d\rho}\left(\mathcal{T}^{\,n-1}{\cal Q}\right) =
\mathcal{T}^{\,n-1} \left[ \mathcal{T}_{,\rho} \left\{
(n-1)\mathcal{T}^{-1}{\cal Q} +1-n(1+c_s^2) \right\} -n(\rho+p)\mathcal{T}_{,\rho\rho} \right],
\label{eq:Q_derivative_Lmrho}
\end{equation}
with $\mathcal{T}_{,\rho\rho} = 2\left[ 1+3c_s^4+3p(c_s^2)_{,\rho}
\right]$. These expressions are local and algebraic, and therefore remain valid in any spacetime, including static, spherically symmetric configurations.

\section{\texorpdfstring{$1+1+2$}{1+1+2} covariant formalism for stars in EMSG}
\label{sec:112_formalism_TOV_EMSG}

The \(1+1+2\) covariant formalism is obtained by further decomposing the standard \(1+3\) covariant approach with respect to a preferred spatial
direction~\cite{Clarkson:2007}. For static and spherically symmetric configurations, the relevant geometries are locally rotationally symmetric
spacetimes of class II (LRS-II), for which all \(1+1+2\) vectors and tensors vanish and the dynamics is encoded entirely in scalar variables. The derivation of the covariant stellar equations in GR, as well as their qualitative phase-space interpretation, was given in Refs.~\cite{Carloni:2018,Naidu:2021} and, in the form used as a benchmark in the present work, in Ref.~\cite{Bittencourt:GRletter}. Here we recall only the minimal geometrical structure needed to extend that construction to EMSG. Related applications of the same covariant strategy to static spherically symmetric solutions in modified gravity can be found in Refs.~\cite{Nzioki:2010,Campbell:2025}.

We introduce a timelike unit vector \(u^a\), chosen as the four-velocity of the static observers comoving with the fluid, and a spacelike unit vector \(e^a\) orthogonal to \(u^a\), satisfying $u_a e^a=0$, and $e_a e^a=1$. If \(h_{ab}=g_{ab}+u_a u_b\) is the projector onto the instantaneous rest space of \(u^a\), the metric on the two-dimensional sheet orthogonal to both \(u^a\) and \(e^a\) is $N_{ab}=h_{ab}-e_a e_b$. In the static LRS-II case, the relevant kinematical and geometrical scalars are the radial acceleration of the static observers ($\mathcal{A} \equiv e^a\dot u_a$), the expansion of the two-dimensional sheets ($\phi \equiv \delta_a e^a$), and the electric part of the Weyl tensor ($\mathcal{E} \equiv C_{abcd}u^a u^c e^b e^d$). Here an overdot denotes covariant differentiation along \(u^a\), while \(\delta_a\) is the covariant derivative projected onto the two-sheet. Since the configurations considered here are static, all time derivatives vanish and the only non-trivial derivatives are along \(e^a\). We denote them by $\hat\psi \equiv e^a D_a\psi$.

For an isotropic static source, the energy-momentum tensor compatible with LRS-II symmetry can be written as
\begin{equation}
    \label{eq:ener_mom_LRS_isot}
    T_{ab}^{({\rm tot})}  = \rho_{\rm tot}u_a u_b + p_{\rm tot}(e_a e_b+N_{ab}),
\end{equation}
where \(\rho_{\rm tot}\) and \(p_{\rm tot}\) denote the total energy density and pressure entering the effective Einstein equations. The minimal set of static LRS-II propagation equations may then be written as \cite{Carloni:2018}
\begin{align}
\hat{\phi} &= -\frac{1}{2}\phi^2+\mathcal{A}\phi-\rho_{\rm tot}-p_{\rm tot},\label{eq:phi_hat_general}\\
\hat p_{\rm tot} &= -(\rho_{\rm tot}+p_{\rm tot})\mathcal{A},\label{eq:pressure_hat_general}\\
\hat{\mathcal{A}} &= -(\mathcal{A}+\phi)\mathcal{A} +\frac{1}{2}(\rho_{\rm tot}+3p_{\rm tot}),\label{eq:A_hat_general}
\end{align}
while $\mathcal{E}$ is determined by the constraint
\begin{equation}
    \mathcal{E}  = -\mathcal{A}\phi +\frac{1}{3}(\rho_{\rm tot}+3p_{\rm tot}) .
\label{eq:constraint_general}
\end{equation}
Equations~\eqref{eq:phi_hat_general}--\eqref{eq:A_hat_general} are purely geometrical once the source has been written in the effective perfect-fluid form~\eqref{eq:ener_mom_LRS_isot}. In GR, they are equivalent to the standard TOV equations in Schwarzschild coordinates~\cite{Carloni:2018}. In modified gravity, the same rewriting is possible whenever the field equations can be cast into Einstein form with an effective source. However, the interpretation of \(\rho_{\rm tot}\) and \(p_{\rm tot}\) must then be theory-dependent. In particular, in theories with curvature corrections one may obtain effective anisotropic or multi-fluid sources~\cite{Campbell:2025}; in EMSG, by contrast, a physical perfect fluid gives rise again to an effective perfect fluid.

Indeed, in the EMSG model considered in Sec.~\ref{sec:theory_EMSG}, the field equations can be written in the GR-like form (\ref{eq:GR_like_EMSG}) and the self-interacting matter corrections preserve the perfect-fluid structure of \(T^{({\rm eff})}_{ab}\). Therefore, in Eq.~\eqref{eq:ener_mom_LRS_isot} we identify
\begin{equation}
\rho_{\rm tot}=\rho_{\rm eff}(\rho,p),
\qquad
p_{\rm tot}=p_{\rm eff}(\rho,p),
\end{equation}
where \(\rho_{\rm eff}\) and \(p_{\rm eff}\) are given by Eqs.~\eqref{eq:rho_eff_general_Lag_p}--\eqref{eq:p_eff_general_Lag_p} for
\(\mathcal{L}_m=p\), and by Eqs.~\eqref{eq:rho_eff_general_Lag_rho}--\eqref{eq:p_eff_general_Lag_rho} for \(\mathcal{L}_m=-\rho\). The covariant stellar equations in EMSG are then
\begin{align}
\hat{\phi} &= -\frac{1}{2}\phi^2+\mathcal{A}\phi-\rho_{\rm eff}-p_{\rm eff},\label{eq:phi_hat_EMSG}\\
\hat p_{\rm eff} &= -(\rho_{\rm eff}+p_{\rm eff})\mathcal{A},\label{eq:pressure_hat_EMSG}\\
\hat{\mathcal{A}} &= -(\mathcal{A}+\phi)\mathcal{A} +\frac{1}{2}(\rho_{\rm eff}+3p_{\rm eff}).\label{eq:A_hat_EMSG}
\end{align}
These are the \textit{covariant TOV equations for EMSG}. They are formally identical to the GR equations, but the variables that gravitate are the effective quantities \(\rho_{\rm eff}\) and \(p_{\rm eff}\), not the physical variables \(\rho\) and \(p\) separately. The physical equation of state \(p=p(\rho)\), together with the algebraic map $(\rho,p)\longmapsto(\rho_{\rm eff},p_{\rm eff})$, therefore determines a closed dynamical system as a GR-like effective stellar model. Equivalently, for fixed \((\eta,n)\), the same map determines the effective equation-of-state parameter and effective sound speed. 

Following Refs.~\cite{Carloni:2018,Naidu:2021,Campbell:2025}, it is useful to introduce a dimensionless radial parameter \(\varrho\) and normalized variables, as follows
\[
\hat X=\phi X', \qquad X'\equiv\frac{dX}{d\varrho}, \qquad \varrho\equiv 2\ln\frac{r}{r_0},
\]
where \(r_0\) is a reference radius. The normalized variables are
\begin{equation}
\Xi=\frac{\phi'}{\phi},
\qquad
Y=\frac{\mathcal{A}}{\phi},
\qquad
\tilde K=\frac{K}{\phi^2},
\qquad
\tilde\mu=\frac{\rho_{\rm eff}}{\phi^2},
\qquad
\tilde p=\frac{p_{\rm eff}}{\phi^2},
\label{eq:normalized_variables_EMSG}
\end{equation}
where \(K\) is the Gaussian curvature of the 2-sheets, introduced for later convenience. In terms of these variables, the dynamical system equivalent to Eqs.~\eqref{eq:phi_hat_EMSG}--\eqref{eq:A_hat_EMSG} becomes
\begin{align}
\tilde K' &= -\tilde K(1+2\Xi),\label{eq:Kprime_preconstraint}\\
Y' &= -Y(1+\Xi+Y) +\frac{1}{2}(\tilde\mu+3\tilde p),\label{eq:Yprime_preconstraint}\\
\tilde p' &= -(\tilde\mu+\tilde p)Y-2\Xi\tilde p,\label{eq:pprime_preconstraint}
\end{align}
with the  corresponding algebraic constraints
\begin{align}
Y&=\tilde K+\tilde p-\frac{1}{4},\label{TOVconstraint1}\\
\Xi&=\tilde K-\tilde\mu-\frac{3}{4}.\label{TOVconstraint2}
\end{align}
Using these constraints to eliminate \(Y\) and \(\Xi\), one obtains the reduced system
\begin{align}
\tilde K' &= -2\tilde K \left(\tilde K-\tilde\mu-\frac{1}{4}\right),\label{eq:Kp_K}\\
\tilde p' &= -\tilde p^2+\tilde p \left(\tilde\mu-3\tilde K+\frac{7}{4}\right) -\tilde\mu \left(\tilde K-\frac{1}{4}\right).\label{eq:Kp_p}
\end{align}
This is the key normalized form of the EMSG stellar equations. If the physical EoS and the EMSG map imply a closed effective relation
of the form $\tilde\mu=\tilde\mu(\tilde p)$, then Eqs.~\eqref{eq:Kp_K}--\eqref{eq:Kp_p} define a two-dimensional autonomous system on the \((\tilde K,\tilde p)\)-plane. If no such effective closure exists, the reduction to a planar system fails, and one must return to a higher-dimensional system involving the original covariant variables or an enlarged set of normalized variables.

This distinction is central to the comparison with GR. In the GR benchmark \cite{Bittencourt:GRletter}, a linear homogeneous equation of state gives the exact closure \(\tilde\mu=\lambda\tilde p\), and the corresponding phase-space structure is organized by invariant sets, finite critical points, and asymptotic fixed points at infinity~\cite{Collins1977,Collins1985,NilssonUgglaLinear,NilssonUgglaPolytropic,HeinzleRohrUggla,WainwrightEllis}. In the present EMSG case, the same geometric vector field is retained, but its closure is controlled by the algebraic transformation from physical to effective variables. Therefore, the relevant question is whether EMSG produces a phase portrait that is globally equivalent to GR, only sectorwise equivalent, asymptotically equivalent, or even genuinely higher-dimensional.

\section{Effective closure and phase-space diagnostics}
\label{sec:qualitative_Kp_effective}

Dynamical-systems methods have also been employed in generalized versions of EMSG, especially in cosmological settings, where they provide a global picture of the asymptotic structure of the model \cite{Bahamonde:2019}. In the present work, we adapt this viewpoint to an effective stellar system ($K>0$) and also consider cases in which the 2-sheets are non-compact ($K\leq0$), for the sake of completeness. 

In this way, consider the autonomous planar system given by Eqs.\ (\ref{eq:Kp_K})-(\ref{eq:Kp_p}), closed by an implicit normalized effective barotropic EoS
\begin{equation}
\tilde\mu=\tilde\mu(\tilde p),\qquad \tilde\mu\in \mathcal{C}^1,\qquad 
\tilde\mu_{\tilde p}:=\frac{d\tilde\mu}{d\tilde p}.
\label{eq:EOS_mup}
\end{equation}
In contrast to ordinary matter, here $\tilde p$ and $\tilde\mu$ may take either sign, as allowed by the nonlinear corrections of EMSG. Therefore, no positivity restriction upon $\tilde p$, $\tilde\mu$ or combinations thereof should be imposed in the phase-plane analysis \textit{a priori}. When the global picture has been established, physical restrictions may be applied.

\subsection{Vector field and nullclines}
\label{subsec:Kp_nullclines}

For the identification of invariant sets and critical regions, let $f(\tilde K,\tilde p)=\tilde K'$ and $g(\tilde K,\tilde p)=\tilde p'$ define the components of a vector field on the $(\tilde K,\tilde p)$-plane. Using Eq.\ \eqref{eq:EOS_mup}, we can write
\begin{align}
f(\tilde K,\tilde p)&=-2\tilde K\Bigl(\tilde K-\tilde\mu(\tilde p)-\frac14\Bigr),\\
g(\tilde K,\tilde p)&=-\tilde p^2+\tilde p\Bigl(\tilde\mu(\tilde p)-3\tilde K+\frac74\Bigr) -\tilde\mu(\tilde p)\Bigl(\tilde K-\frac14\Bigr).
\end{align}
According to the vanishing of each component separately, we have the nullclines organized as:

\paragraph{$\tilde K$-nullclines.} From Eq.\ \eqref{eq:Kp_K}, we get
\begin{equation}
\mathcal N_{\tilde K}=\{\tilde K=0\}\ \cup\ \Bigl\{\tilde K=\tilde\mu+\frac14\Bigr\}.
\label{eq:K_nullclines}
\end{equation}

\paragraph{$\tilde p$-nullcline.} A key structural feature is that Eq.\ \eqref{eq:Kp_p} is affine in $\tilde K$. Thus, it is possible to write as
\begin{equation}
g(\tilde K,\tilde p)=-(3\tilde p+\tilde\mu)\,\tilde K + q(\tilde p),
\quad \mbox{with}\quad
q(\tilde p):=-\tilde p^2+\tilde p\Bigl(\tilde\mu+\frac74\Bigr)+\frac14\,\tilde\mu.
\label{eq:g_affine_in_K}
\end{equation}
Hence, whenever $3\tilde p+\tilde\mu(\tilde p)\neq 0$, the $\tilde p$-nullcline is
\begin{equation}
\mathcal N_{\tilde p}=\left\{
\tilde K_{\tilde p}(\tilde p)\equiv\tilde K=\frac{q(\tilde p)}{3\tilde p+\tilde\mu(\tilde p)}\right\}.
\label{eq:p_nullcline}
\end{equation}
In the degenerate case $3\tilde p+\tilde\mu(\tilde p)=0$, one has $g(\tilde K,\tilde p)=q(\tilde p)$, so $\tilde p'=0$ holds \emph{for all $\tilde K$} if and only if simultaneously
\begin{equation}
3\tilde p+\tilde\mu(\tilde p)=0 \qquad\text{and}\qquad q(\tilde p)=0,
\label{eq:vertical_invariant_condition}
\end{equation}
in which case the line $\tilde p=\tilde p_0$ is an invariant set.

From this, we can rewrite Eq.\ \eqref{eq:g_affine_in_K} as
\begin{equation}
\tilde p' = -(3\tilde p+\tilde\mu)\,\bigl(\tilde K-\tilde K_{\tilde p}(\tilde p)\bigr).
\label{eq:pprime_sign_rule}
\end{equation}
So the sign of $\tilde p'$ is determined by the relative position to the curve $\tilde K=\tilde K_{\tilde p}(\tilde p)$, modulated by the sign of $3\tilde p+ \tilde\mu$. Similarly, from Eq.\ \eqref{eq:Kp_K}, we obtain
\begin{equation}
\tilde K' = -2\tilde K\Bigl(\tilde K-\tilde\mu(\tilde p)-\frac14\Bigr)
\quad\Rightarrow\quad
\tilde K'\gtrless 0\ \Longleftrightarrow\
\tilde K\lessgtr \tilde\mu(\tilde p)+\frac14\ \ \text{if}\ \ \tilde K>0,
\label{eq:Kprime_sign_rule}
\end{equation}
with the inequalities reversed when $\tilde K<0$.

Because $\tilde K'$ has an overall factor $\tilde K$, we have
\begin{equation}
\tilde K(\varrho_0)=0\ \Longrightarrow\ \tilde K(\varrho)\equiv 0,\quad\forall\,\varrho.
\label{eq:K0_invariant}
\end{equation}
Therefore, the half-planes $\tilde K>0$ and $\tilde K<0$ are forward invariant, as well as the line $\tilde K=0$. As $\tilde{K}$ is a rescaling of the Gaussian curvature of the 2-sheet, this feature means that topological transitions are not allowed for the 2-sheet. On the other hand, the curve $\tilde K=\tilde\mu(\tilde p)+\tfrac14$ is not invariant generically, but only in the special situations: $\tilde\mu_{\tilde p}=0$ locally; or $\tilde p'=0$ along the curve, which typically happens only at equilibria.

If a value $\tilde p_0$ satisfies the pair of Eqs.\ \eqref{eq:vertical_invariant_condition}, then $\tilde p'=0$ for all $\tilde K$
on $\tilde p=\tilde p_0$, and the reduced dynamics on that invariant line is the scalar Riccati-type equation
\begin{equation}
\tilde K'=-2\tilde K\Bigl(\tilde K-\tilde\mu(\tilde p_0)-\frac14\Bigr).
\end{equation}
Such invariant lines represent ``frozen-pressure'' sectors, namely, the pressure achieves a constant value for asymptotic values of the radial parameter, specific to the chosen EoS.

\subsection{Fixed points and linear stability}
\label{subsec:Kp_equilibria}

Equilibria $(\tilde K_\star,\tilde p_\star)$ satisfy $f(\tilde K_\star,\tilde p_\star)=0=g(\tilde K_\star,\tilde p_\star)$. From Eq.\ \eqref{eq:K_nullclines}, we see that they must lie on one of the two $\tilde K$-nullclines, namely, on their intersections with the $\tilde p$-nullclines. For simplicity, we separate them by

\paragraph{Branch I: $\tilde K_\star=0$.}
Setting $\tilde K=0$ in Eq.\ \eqref{eq:g_affine_in_K} gives $\tilde p'=q(\tilde p)$, hence
\begin{equation}
(\tilde K_\star,\tilde p_\star)=(0,\tilde p_\star)
\quad\text{with}\quad
q(\tilde p_\star)=0.
\label{eq:equilibria_branchI}
\end{equation}
This is an implicit scalar equation for $\tilde p_\star$ determined by the EoS. The number of real roots of this equation corresponds to the number of equilibrium points on this branch. There is also room for degenerate equilibrium lines, if $q(\tilde p)\equiv 0$ on an interval. Then the whole segment $\{\tilde K=0,\ \tilde p\in I\}$ will consist of equilibria. 

\paragraph{Branch II: $\tilde K_\star=\tilde\mu(\tilde p_\star)+\tfrac14$.}
Substituting $\tilde K=\tilde\mu(\tilde p)+\tfrac14$ into Eq.\ \eqref{eq:Kp_p} yields the reduced equilibrium condition
\begin{equation}
\tilde p_\star^2-\tilde p_\star\bigl(1-2\tilde\mu(\tilde p_\star)\bigr)+\tilde\mu(\tilde p_\star)^2=0.
\label{eq:equilibria_branchII}
\end{equation}
Again, this is an implicit condition involving the EoS curve $\tilde\mu=\tilde\mu(\tilde p)$.

The Jacobian of the vector field is
\begin{equation}
J(\tilde K,\tilde p)=
\begin{pmatrix}
\displaystyle \frac{\partial f}{\partial \tilde K} &
\displaystyle \frac{\partial f}{\partial \tilde p}\\[2mm]
\displaystyle \frac{\partial g}{\partial \tilde K} &
\displaystyle \frac{\partial g}{\partial \tilde p}
\end{pmatrix}
=
\begin{pmatrix}
2\Bigl(\tilde\mu+\frac14-2\tilde K\Bigr) & 2\tilde K\,\tilde\mu_{\tilde p}\\[1mm]
-(3\tilde p+\tilde\mu) &
-2\tilde p+\tilde\mu-3\tilde K+\frac74 + \tilde\mu_{\tilde p}\Bigl(\tilde p-\tilde K+\frac14\Bigr)
\end{pmatrix}.
\label{eq:Jacobian_general_effective}
\end{equation}
For an equilibrium $(\tilde K_\star,\tilde p_\star)$, stability is determined by the eigenvalues ($l_1$, $l_2$) of $J_\star$,
or equivalently by its trace $\tau$ and determinant $\Delta$. Thus, $\Delta<0$ indicates it is a saddle; for $\Delta>0,\ \tau<0$ it is a stable node; and for $\Delta>0,\ \tau>0$ we have an unstable node. A focus occurs when $\tau^2-4\Delta<0$.

The stability on Branch I is obtained from the points $(0,\tilde p_\star)$ satisfying Eq.\ \eqref{eq:equilibria_branchI}, for which $J_\star$ is lower triangular. Hence,
\begin{equation}
l_1=2\Bigl(\tilde\mu_\star+\frac14\Bigr),\qquad
l_2\equiv q_{\tilde p,\star}=-2\tilde p_\star+\tilde\mu_\star+\frac74+\tilde\mu_{\tilde p,\star}\Bigl(\tilde p_\star+\frac14\Bigr),
\label{eq:eigs_branchI_effective}
\end{equation}
where $\tilde\mu_\star:=\tilde\mu(\tilde p_\star)$, $\tilde\mu_{\tilde p,\star}:=\tilde\mu_{\tilde p}(\tilde p_\star)$, and $q_{\tilde p,\star}=(dq/d\tilde p)|_{\tilde p=\tilde p_\star}$.
Thus, depending on the EoS, equilibria on $\tilde K=0$ may be sinks, sources, or saddles; in particular, the signs of
$\tilde\mu_\star+\tfrac14$ and $q_{\tilde p,\star}$ are decisive for the $\tilde K$-direction.

At a Branch II equilibrium, substituting $\tilde K_\star=\tilde\mu_\star+\tfrac14$ into Eq.\ \eqref{eq:Jacobian_general_effective}
gives
\begin{equation}
J_\star=
\begin{pmatrix}
-2\Bigl(\tilde\mu_\star+\frac14\Bigr) 
& 2\Bigl(\tilde\mu_\star+\frac14\Bigr)\tilde\mu_{\tilde p,\star}\\[1mm]
-(3\tilde p_\star+\tilde\mu_\star) &
1-2(\tilde p_\star+\tilde\mu_\star)+\tilde\mu_{\tilde p,\star}(\tilde p_\star-\tilde\mu_\star)
\end{pmatrix}.
\label{eq:Jacobian_branchII_effective}
\end{equation}
Consequently,
\begin{align}
\tau &= \frac12-2\tilde p_\star-4\tilde\mu_\star+\tilde\mu_{\tilde p,\star}(\tilde p_\star-\tilde\mu_\star),
\label{eq:trace_branchII_effective}\\
\Delta &
= 2\Bigl(\tilde\mu_\star+\frac14\Bigr)\Bigl[-1+2(\tilde p_\star+\tilde\mu_\star)\bigl(1+\tilde\mu_{\tilde p,\star}\bigr)\Bigr].
\label{eq:det_branchII_effective}
\end{align}
These expressions provide an EoS-dependent classification of the Branch II equilibria.

Given that the system of equations corresponding to the TOV equations (\ref{eq:phi_hat_EMSG})-(\ref{eq:A_hat_EMSG}) could be reduced to an autonomous planar dynamical system (\ref{eq:Kp_K})-(\ref{eq:Kp_p}), without loss of the physical character of the dynamical variables, yet its qualitative analysis remains rather general unless an EoS is specified. We shall take a step further and analyze what EMSG would look like when a linear EoS is assumed for the physical variables $\rho$ and $p$. This type of equation is highly idealized, but it is expected to correspond to the first-order term in a power-series expansion of more realistic EoSes.

\section{Linear physical EoS}
\label{sec:phase_portraits_linearEOS}

In this section, we specialize the fundamental autonomous dynamical system (\ref{eq:Kp_K})-(\ref{eq:Kp_p}) to the case in which the physical energy density and pressure obey
\begin{equation}
\rho=\lambda p,\qquad \lambda=\mathrm{const.},\qquad
c_s^2=\frac{dp}{d\rho}=\frac{1}{\lambda}\quad (\lambda\neq 0).
\label{eq:physical_linearEOS}
\end{equation}
Therefore, a physically reasonable assumption is $\lambda\notin(-1,1)$. The effective quantities $\rho_{\rm eff}(\rho,p)$ and $p_{\rm eff}(\rho,p)$ are given by
Eqs.~(\ref{eq:rho_eff_general_Lag_p})-(\ref{eq:p_eff_general_Lag_p}) for $\mathcal{L}_m=p$, and by Eqs.~(\ref{eq:rho_eff_general_Lag_rho})-(\ref{eq:p_eff_general_Lag_rho}) for $\mathcal{L}_m=-\rho$.
As mentioned before, since $\tilde p$ and $\tilde\mu$ are effective variables, we do not impose sign restrictions on them. Thus, the analysis is performed on the full $(\tilde K,\tilde p)$ plane, as long as the vector field is well-defined.

\subsection{GR benchmark revisited}
\label{subsec:GR_benchmark_SecVI}

Several EMSG sectors below reduce to the same normalized vector field as the GR linear-EoS problem, possibly with a different or sector-dependent effective slope. For this reason, we briefly recall the GR benchmark obtained in Ref.~\cite{Bittencourt:GRletter}.

In the GR case, the effective and physical variables coincide, such that the linear physical EoS \eqref{eq:physical_linearEOS} is the exact normalized closure $\tilde\mu=\lambda\,\tilde p$. Substitution of this expression into Eqs.~\eqref{eq:Kp_K}--\eqref{eq:Kp_p} gives the GR polynomial planar system
\begin{align}
\tilde K' &= -2\tilde K \left( \tilde K-\lambda\tilde p-\frac{1}{4} \right),\label{eq:GR_K}\\
\tilde p' &= \tilde p \left[(\lambda-1)\tilde p -(\lambda+3)\tilde K +\frac{\lambda+7}{4}\right].\label{eq:GR_p}
\end{align}

\begin{table}[ht]
\centering
\begin{tabular}{|c|c|c|c|c|c|}
\hline
{\bf Fixed point}
&
\boldmath{$\lambda<-7$}
&
\boldmath{$-7<\lambda<\lambda_-$}
&
\boldmath{$\lambda_-<\lambda<\lambda_+$}
&
\boldmath{$\lambda_+<\lambda<1$}
&
\boldmath{$\lambda>1$}
\\
\hline
$P_0$       & saddle & source & source & source & source \\
$P_{1/4}$   & saddle & saddle & saddle & saddle & saddle \\
$P_{K0}$    & source & saddle & sink   & saddle & sink   \\
$P_{\rm int}$ & sink & sink & saddle & sink & sink \\
\hline
\end{tabular}
\caption{Stability of the finite fixed points of Eqs. (\ref{eq:GR_K})-(\ref{eq:GR_p}): $P_0=(0,0)$, $P_{1/4} =\left(\frac14,0\right)$, $P_{K0}=\left(0,-\frac{7+\lambda}{4(\lambda-1)} \right)$, and $P_{\rm int}=\left( \frac14+\frac{\lambda}{(1+\lambda)^2}, \frac{1}{(1+\lambda)^2} \right)$. The values $\lambda=-7$ and $\lambda_{\pm}=-3\pm2\sqrt2$ correspond to non-hyperbolic cases. The values $\lambda=-1$ and $\lambda=1$ require separate treatment because some of the equilibrium-point expressions become singular.}
\label{tab:stability_equil}
\end{table}

As already discussed in Ref.~\cite{Bittencourt:GRletter}, it admits four fixed points, whose coordinates and stabilities are summarized in Table \ref{tab:stability_equil}. For ordinary GR stellar models, one usually restricts attention to the physical sector in which the density and pressure are non-negative, enforcing $\lambda>0$. Moreover, if causal restrictions are applied, it requires $\lambda\geq1$ in order to have $c_s^2<1$, eliminating, for instance, $P_{K0}$ from the physical sector. In the normalized variables, it will be translated as $\tilde K>0$, $\tilde p\geq0$, and $\tilde\mu\geq0$. 

For the asymptotic structure, we shall not repeat the compactification analysis in detail. We only recall the result needed below. Whenever the effective closure is linear or asymptotically linear, with
\begin{equation}
\lambda_\infty:=\lim_{|\tilde p|\to\infty}\frac{\tilde\mu(\tilde p)}{\tilde p},
\end{equation}
the leading homogeneous part of the vector field is
\begin{align}
\tilde K'&\sim -2\tilde K^2+2\lambda_\infty \tilde K\tilde p,
\label{eq:asympK}\\
\tilde p'&\sim (\lambda_\infty-1)\tilde p^2-(\lambda_\infty+3)\tilde K\tilde p.
\label{eq:asympp}
\end{align}
The corresponding fixed directions at infinity and their stability are those of the GR benchmark, summarized in Table~\ref{tab:fixed_points_alpha} (see Ref.~\cite{Bittencourt:GRletter} for the compactification details).

\begin{table}[ht]
\centering
\begin{tabular}{|c|c|c|c|}
\hline
\textbf{Invariant direction: eigenvalues} & $\boldsymbol{\lambda_\infty<-1}$ & $\boldsymbol{-1<\lambda_\infty<1}$ & $\boldsymbol{\lambda_\infty>1}$ \\
\hline
$E_{\tilde p=0}:\left(2,\,-\lambda_\infty-1\right)$ 
& Source/Sink & Saddle & Saddle \\
$E_{\tilde K=0}:\left(1-\lambda_\infty,\,\lambda_\infty+1\right)$ 
& Saddle & Source/Sink & Saddle \\
$E_{\tilde p=-\tilde K}:\left(2\lambda_\infty+2,\,\lambda_\infty+1\right)$ 
& Sink/Source & Source/Sink & Source/Sink\\
\hline
\end{tabular}
\caption{Stability of the asymptotic fixed points as a function of $\lambda_\infty$.}
\label{tab:fixed_points_alpha}
\end{table}

However, in the EMSG case, the variables entering the reduced system are the effective ones. Even if the physical variables $\rho$ and $p$ obey the usual positivity requirements, the nonlinear matter corrections may render $\rho_{\rm eff}$ or $p_{\rm eff}$ negative in some region of parameter space. Therefore, in the EMSG phase-space analysis one should not impose the GR positivity restrictions on $\tilde\mu$ and $\tilde p$ a priori. This is precisely why the full $(\tilde K,\tilde p)$ plane, or sectorwise portions of it, may become relevant. 

At the same time, whenever an EMSG model yields an exact linear effective closure $\tilde\mu=\lambda_{\rm eff}\tilde p$, its finite phase-space structure can be obtained from the GR benchmark by the replacement $\lambda\longrightarrow \lambda_{\rm eff}$. If the effective closure is only piecewise linear, the same identification holds by sector. If no relation of the form $\tilde\mu=\tilde\mu(\tilde p)$ exists, the planar reduction is not available and the full covariant system must be used. 

\subsection{EMSG with \texorpdfstring{$\mathcal{L}_m=p$}{}}
\label{subsec:Lmp}
As one would expect, the linear EoS between the physical variables does not automatically imply the same relation for the effective ones in EMSG. Nevertheless, there are interesting cases in which this can be done without recalling the original 3D system.

Let $\bar p:=p/\phi^2$ denote the normalized physical pressure. With $\rho=\lambda p$ and $\mathcal{T}=(\lambda^2+3)\,p^2$, the normalized effective variables $\tilde p=p_{\rm eff}/\phi^2$ and $\tilde\mu=\rho_{\rm eff}/\phi^2$ take the following form in terms of $|\bar p|$:
\begin{align}
\tilde p &= \bar p + \frac{\eta}{2}\,(\lambda^2+3)^{n}\,\phi^{4n-2}\,|\bar p|^{2n},
\label{eq:Lmp_ptilde_param}\\
\tilde\mu &= \lambda \bar p + \eta\,(\lambda^2+3)^{n-1}\Big[\Big(n-\frac12\Big)(\lambda^2+3)+4n\lambda\Big]\,\phi^{4n-2}\,|\bar p|^{2n}.
\label{eq:Lmp_mutilde_param}
\end{align}
Note that the explicit presence of $\phi$ inhibits the construction of a closure relation $\tilde\mu(\tilde p)$ through $|\bar p|$ in general, unless particular cases are taken into account.

For instance, $\phi$ is absent when $n=\frac12$, so that the normalized effective variables become homogeneous of degree one in the normalized physical pressure $\bar p$.  However, the resulting closure must be understood \emph{branchwise}, because it depends on the sign of $\bar p$. Introducing $\sigma:=\operatorname{sgn}\bar p$, Eqs.~(62) and (63) become
\begin{equation}
\tilde p=\Delta_\sigma\,\bar p, \qquad \tilde\mu=\Gamma_\sigma\,\bar p,
\label{eq:Lmp_nhalf_branch_map}
\end{equation}
with $\Delta_\sigma\equiv1+(\sigma\,\eta\,\sqrt{\lambda^2+3})/2$ and $\Gamma_\sigma\equiv\lambda + 2\sigma\eta\lambda/\sqrt{\lambda^2+3}$. Therefore, on each branch $\sigma=\pm1$, provided $\Delta_\sigma\neq0$, one obtains the linear effective closure
\begin{equation}
\tilde\mu=\lambda_\sigma\,\tilde p,
\quad\mbox{with}\quad
\lambda_\sigma:=\frac{\Gamma_\sigma}{\Delta_\sigma}
=
\frac{\lambda \pm \dfrac{2\eta\lambda}{\sqrt{\lambda^2+3}}}
{1\pm \dfrac{\eta}{2}\sqrt{\lambda^2+3}} .
\label{eq:Lmp_nhalf_mu_of_p}
\end{equation}
Hence, the effective dynamics is GR-like on each branch separately, under the replacement $\lambda_{\rm GR}\to\lambda_\sigma$.

It is important to stress, however, that Eq.~\eqref{eq:Lmp_nhalf_mu_of_p} does \emph{not} necessarily define a single-valued global relation $\tilde\mu=\tilde\mu(\tilde p)$ on the whole effective plane. The image of
each physical branch is controlled by the sign of $\Delta_\sigma$.  Only in the weak-coupling regime
\begin{equation}
|\eta|<\frac{2}{\sqrt{\lambda^2+3}},
\label{eq:branch_preserving_condition}
\end{equation}
do both $\Delta_+$ and $\Delta_-$ remain positive, so that the branch $\sigma=+1$ maps to $\tilde p>0$ and the branch $\sigma=-1$ maps to $\tilde p<0$.  In that case, the two half-planes of the $(\tilde K,\tilde p)$-plane are directly associated with the two physical branches, and the system is genuinely piecewise linear in $\tilde p$. Thus, it will be a slight deformation of the GR case analyzed in Ref.\ \cite{Bittencourt:GRletter} 

In the phase-space analysis below, we use the effective variables as the primary dynamical variables.  Thus \(\tilde p\) is regarded as the coordinate of the reduced autonomous system and is allowed to run over the whole real line. The two half-planes \(\tilde p>0\) and \(\tilde p<0\) should therefore be understood as effective sectors of the piecewise-linear closure, not necessarily as the direct images of the physical branches \(\bar p>0\) and
\(\bar p<0\). In the branch-preserving regime
\eqref{eq:branch_preserving_condition}, these notions coincide.  For stronger couplings, however, one of the factors \(\Delta_\sigma\) may change sign, so that the map \(\bar p\mapsto\tilde p\) reverses orientation on that branch. Consequently, the physical interpretation of a given orbit must be recovered only after mapping the effective variables back to the physical variables,
whereas the full \((\tilde K,\tilde p)\)-plane remains useful for organizing the global effective dynamics.

\begin{figure}
    \centering
 \includegraphics[width=7.5cm,height=7.5cm]{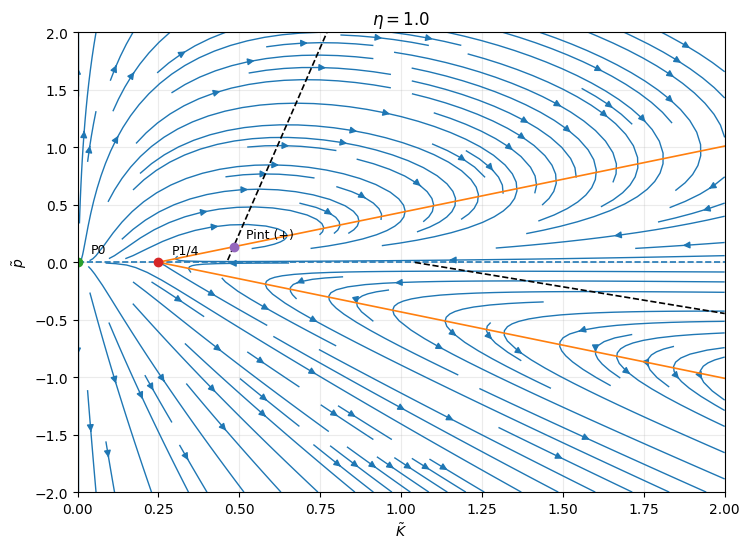}
    \includegraphics[width=7.5cm,height=7.5cm]{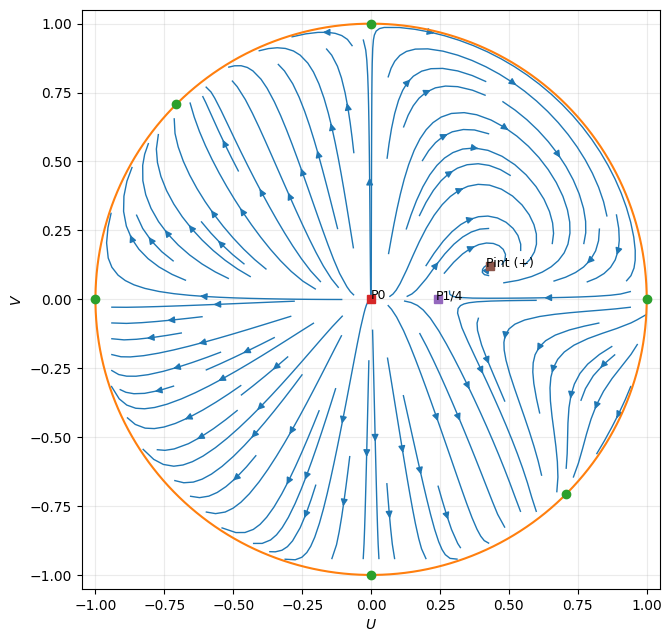}
    \includegraphics[width=7.5cm,height=7.5cm]{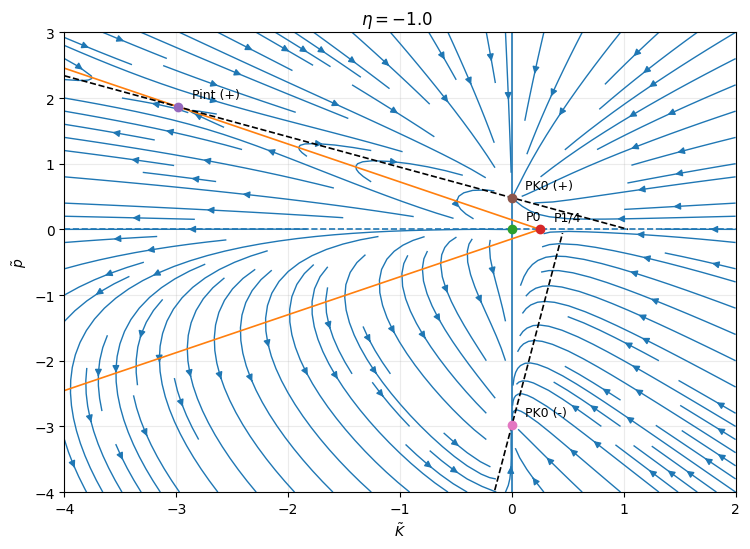}
    \includegraphics[width=7.5cm,height=7.5cm]{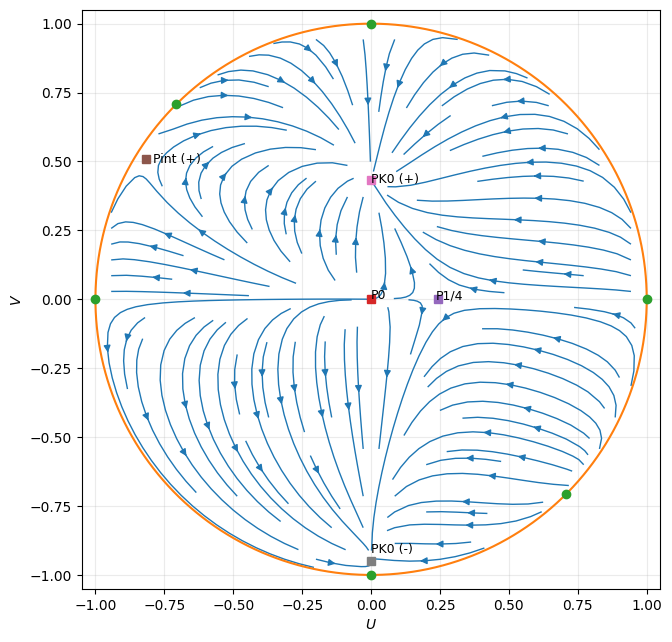}
\caption{Radiation case in EMSG with \(L_m=p\) and \(n=\tfrac12\). For this value of \(n\), the effective closure is defined branchwise according to \(\sigma\). On each branch, the system is topologically equivalent to the GR linear-EoS case under the replacement \(\lambda_{\rm GR}\to\lambda_\sigma\), with \(\lambda_\sigma\) given by Eq.~\eqref{eq:Lmp_nhalf_mu_of_p}; consequently, the finite fixed points and the asymptotic directions follow the GR classification, with the stability at infinity determined by Tab.~\ref{tab:fixed_points_alpha}. The half-planes \(\tilde p>0\) and \(\tilde p<0\) are effective sectors of this reduced system and need not coincide with the images of the physical branches \(\bar p>0\) and \(\bar p<0\) when the branch-preserving condition is violated. Top panels: \(\eta=1\).  In this case, the image of one branch does not contain the GR-type equilibrium \(P_{K0}\), so along the nullcline \(\tilde K=0\) only the origin is represented.  Bottom panels: \(\eta=-1\).  The extended branch structure yields the four sectorwise GR-type finite critical points together with an additional saddle in the quadrant \(\tilde p>0\), \(\tilde K<0\).  Thus, the change in the number of finite equilibrium points between the two cases is a genuine consequence of the nonlinear EMSG branch map, rather than a local bifurcation of the GR polynomial vector field.  The right panels show the corresponding Poincar\'e compactifications of the extended branch portraits under the map $\{U,V\}=(1+\tilde K^2+\tilde p^2)^{-\frac12}\{\tilde K,\tilde p\}$.
}
\label{fig:Lp_n12_rad}
\end{figure}

For the sake of illustration, Fig.~\ref{fig:Lp_n12_rad} shows the radiation
case \(\lambda=3\) for representative values \(\eta=\pm1\). Since the branch-preserving threshold is then \(|\eta|=1/\sqrt{3}\), both choices lie outside the regime \eqref{eq:branch_preserving_condition}. The figure should therefore be read as the phase portrait of the effective piecewise-linear system in the \((\tilde K,\tilde p)\)-plane. On each effective sector, the vector field is GR-like under the replacement \(\lambda_{\rm GR}\to\lambda_\sigma\), but the half-planes \(\tilde p>0\) and \(\tilde p<0\) should not be identified, in general, with positive and negative physical pressure. This is why changing \(\eta\) can change the number of GR-type equilibrium points represented in the effective portrait, while the physical stellar sector must still be selected separately by imposing the desired physical branch.

From this perspective, the change in the number of equilibrium points is not a contradiction, but a genuine manifestation of the nonlinear EMSG map. On each effective branch, the finite critical points are exactly those of the GR benchmark with $\lambda_{\rm GR}\to\lambda_\sigma$; however, a given GR-type critical point is represented in the effective plane only if its value of $\tilde p$ belongs to the image of that branch. Thus, depending on $\eta$, some critical points may leave the represented sector, while others may enter it. In particular, in the top-left panel of Fig.~\ref{fig:Lp_n12_rad}, there is no equilibrium along the nullcline $\tilde K=0$ besides the origin, meaning that the corresponding GR-type point $P_{K0}$ does not belong to the represented image of that branch.  By contrast, for negative $\eta$, the four sectorwise GR finite points are present, while an additional saddle appears in the quadrant $\tilde p>0$ and $\tilde K<0$ due to the different branch assignment in the extended effective plane.

Therefore, for fixed $\lambda$, the parameter $\eta$ changes the effective slopes $\lambda_\sigma$ and may also alter the way the two physical branches are mapped into the effective plane.  As $\eta\to0^\pm$, one recovers $\lambda_+=\lambda_-=\lambda$, the branch distinction disappears, and the phase portrait deforms continuously to the differentiable GR case.

The effective linear EoS is also valid in the regime where the nonlinear corrections of the EMSG are dominant. If $n>1/2$ and $|\bar p|^{2n-1}\gg1$, the linear terms in Eqs.\ \eqref{eq:Lmp_ptilde_param} and \eqref{eq:Lmp_mutilde_param} can be neglected, and we get an asymptotic linear EoS with effective EoS parameter
\begin{equation}
    \lambda_{\infty}=\frac{2\Big[\Big(n-\frac12\Big)(\lambda^2+3)+4n\lambda\Big]}{\lambda^2+3}.
\end{equation}
This is a remarkable result since it ensures that the 2D system \eqref{eq:asympK}-\eqref{eq:asympp} is still responsible for the dynamics at infinity, independently of $\eta$, for any $\lambda$ and $n>1/2$.

By enforcing Eqs.\ \eqref{eq:Lmp_ptilde_param}-\eqref{eq:Lmp_mutilde_param} to satisfy an exact linear closure relation in the effective variables, we get a cubic polynomial for $\lambda$, whose trivial case is the cosmological-constant-type branch ($\lambda=-1$), and the other possible cases are given by 
\begin{equation}
\lambda_{1,2}=n\pm\sqrt{n^2+6n-3}.
\end{equation}
The argument of the square root vanishes for $n_{\pm}=-3\pm2\sqrt{3}$, therefore, for $n\in(n_-,n_+)$ there is no real $\lambda$ satisfying the equation above.

\begin{figure}
    \centering
    \includegraphics[width=7.5cm,height=7.5cm]{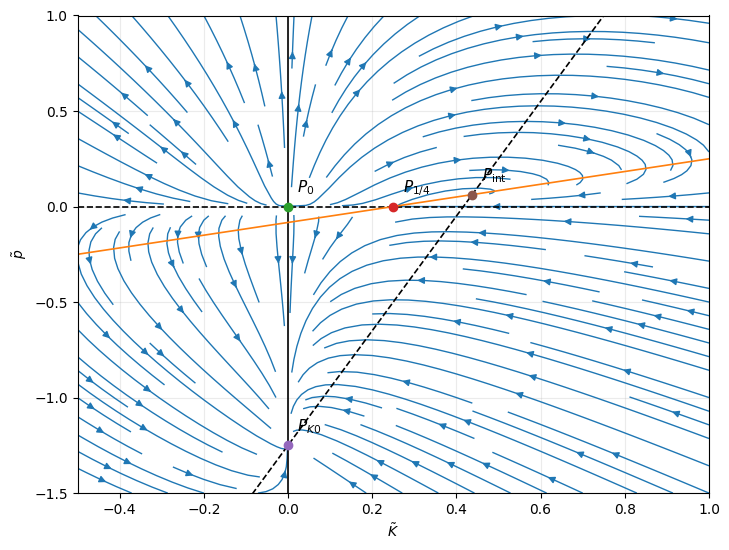}
    \includegraphics[width=7.5cm,height=7.5cm]{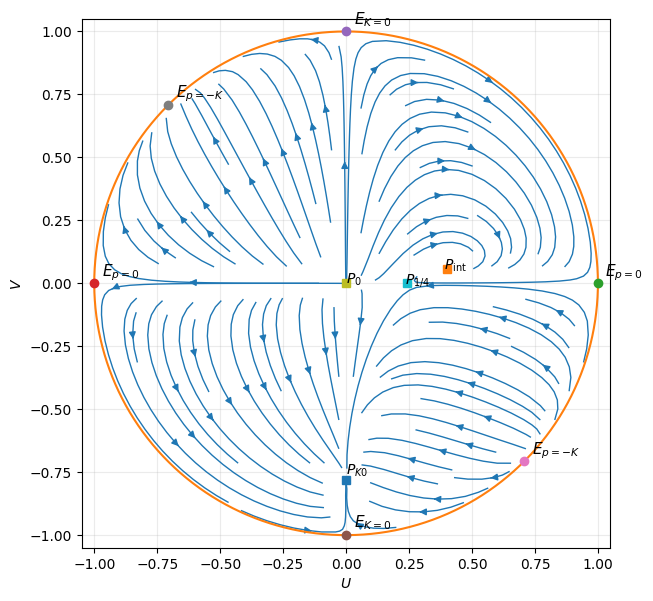}
    \caption{Radiation case in EMSG with \(\mathcal L_m=p\) and \(n=1\). This sector is dynamically equivalent to the GR radiation when written in terms of the effective variables, but the full effective \((\tilde K,\tilde p)\)-plane is displayed because \(\tilde p\) and \(\tilde\mu\) are not restricted a priori. Left panel: nullclines \(\tilde K'=0\) and \(\tilde p'=0\), shown as solid and dashed curves, respectively. Their intersections give \(P_0\), \(P_{1/4}\), \(P_{K0}\), and \(P_{\rm int}\). The point \(P_{K0}\), usually discarded in the positive-pressure GR stellar sector, is retained here because it organizes the full effective EMSG flow. Right panel: Poincar\'e compactification of the same effective system under the map $\{U,V\}=(1+\tilde K^2+\tilde p^2)^{-\frac12}\{\tilde K,\tilde p\}$..}
    \label{fig:Lp_n1_rad}
\end{figure}

In particular, for $n=1$, we obtain $\lambda=3$, recovering the known result that EMSG behaves exactly like GR for relativistic stars, regardless of $\eta$. Obviously, the only difference is the mathematical expressions between the physical and the effective variables, which do not affect the normalized phase portrait. Hence, this case is globally topologically equivalent to the GR radiation benchmark, leading to the same phase portrait depicted in Fig.\ \ref{fig:Lp_n1_rad}. 

\begin{figure}
    \centering
 \includegraphics[width=16cm,height=13cm]{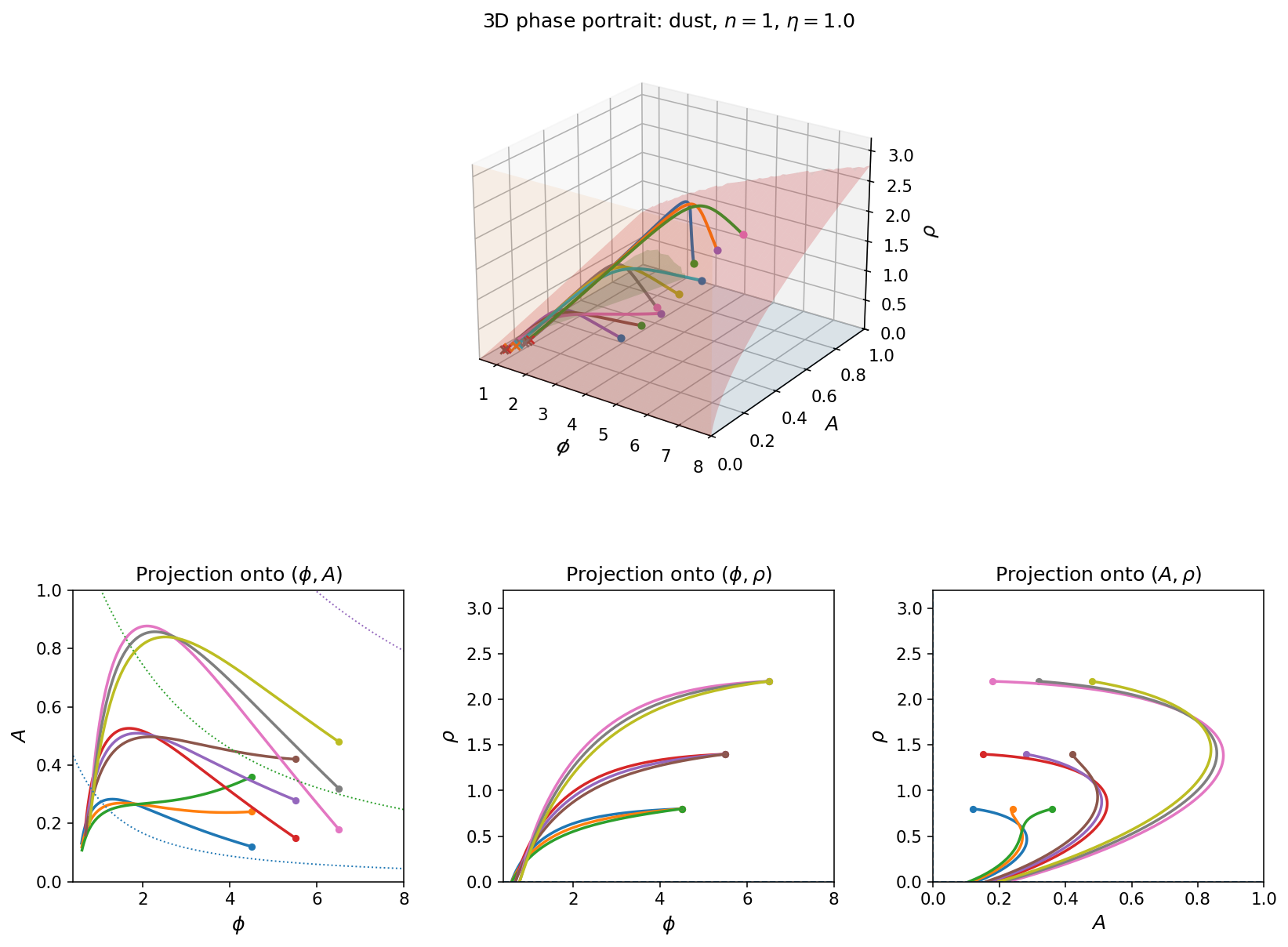}
    \caption{Dust sector in EMSG for $L_m=p$ and $n=1$, with $\eta=1.0$. Top: representative integral curves of the full 3D covariant system in the physically relevant sector $(\phi,\mathcal{A},\rho)>0$, showing orbits that connect a regular central asymptotic regime (indicated by the dots) to the vacuum boundary $\rho=0$. The semitransparent coordinate planes are included only as guides to the eye. Bottom: projections of the same trajectories onto the planes $(\phi,\mathcal{A})$-plane, $(\phi,\rho)$-plane, and $(\mathcal{A},\rho)$-plane, respectively. The flow illustrates that there is no finite equilibrium with $\rho>0$; instead, physically admissible dust configurations are organized as monotonic orbits of the full 3D system.}
    \label{fig:Lp_dust}
\end{figure}

Another interesting case to discuss is the physical dust. Notice that this requires setting $p=0$ directly into Eqs.~\eqref{eq:rho_eff_general_Lag_p}-\eqref{eq:p_eff_general_Lag_p}. Thus, the contributions from EMSG to the effective fluid reduce to
\begin{equation}
\rho_{\rm emsg}=\eta\left(n-\frac12\right)\rho^{2n},
\qquad
p_{\rm emsg}=\frac{\eta}{2}\rho^{2n},
\label{eq:rho_p_emsg_Lp_dust}
\end{equation}
so that EMSG endows pressureless matter with a nonvanishing effective pressure. In this case, the system cannot be reduced to a closed planar dynamics in \((\tilde K,\tilde p)\), since \(\tilde\mu\) cannot be written as a function of \(\tilde p\) solely; one must instead work with the full 3D system \eqref{eq:phi_hat_EMSG}-\eqref{eq:A_hat_EMSG}. For $n=1$, it is possible to show that the only finite critical point is the origin \((\phi=0,\mathcal{A}=0,\rho=0)\), and there is no finite equilibrium with \(\rho>0\). Therefore, a static dust configuration is not organized around an interior fixed point; rather, it must be understood as an orbit of the full 3D flow connecting a regular central asymptotic regime to the vacuum surface $\rho=0$ (see Fig.\ \ref{fig:Lp_dust}).

The physically relevant branch is \(\eta>0\). In the sector \(\phi>0\), \(A>0\), \(\rho>0\), one has
\[
\hat{\rho}=-\frac{1+\eta\rho}{\eta}\,A<0,
\]
so the density decreases monotonically along the flow, ruling out periodic or recurrent behavior and implying that the stellar surface is reached at \(\rho=0\), equivalently \(p_{\rm eff}(R)=0\). Moreover, the effective EoS given by Eq.\ \eqref{eq:bar_w_Lp} and sound speed given by Eq.\ \eqref{eq:bar_cs2L_p} reduce to
\begin{equation}
\label{eq:Lp_dust_eff_w_cs2}
w_{\rm eff}=\frac{\eta\rho}{2+\eta\rho},
\qquad
[c_s^2]_{\rm eff}=\frac{\eta\rho}{1+\eta\rho},
\end{equation}
satisfying \(0\leq w_{\rm eff}<1\) and \(0\leq c_{\rm eff}^{2}<1\), so dust behaves as an effectively regular fluid with increasing stiffness at high density. Regarding the other variables, it is expected that $\phi\rightarrow+\infty$ and $\mathcal{A}=0$ at the center, but reach nonzero finite values at the star surface (see Sec.\ \ref{sec:metric_TOV_map}). By contrast, the branch \(\eta<0\) is qualitatively pathological: \(c_{\rm eff}^{2}\) becomes negative or singular, and \(\hat{\rho}\) may become positive, so the density can grow outward. Finally, although the origin is nonhyperbolic for \(\eta>0\), it is not Lyapunov stable in the full phase space (see the subspace $\mathcal{A}=0=\rho$); hence it should be interpreted only as a boundary vacuum point, not as an attractor for the physical stellar flow.

\subsection{EMSG with \texorpdfstring{$\mathcal{L}_m=-\rho$}{}}
\label{subsec:LmminusRho}

We shall proceed similarly to the previous case when $\mathcal{L}_m=-\rho$. By using \eqref{eq:physical_linearEOS} and the normalized pressure $\bar p$ in Eqs.\ \eqref{eq:rho_eff_general_Lag_rho}-\eqref{eq:p_eff_general_Lag_rho}, the normalized effective energy density and pressure become, respectively,
\begin{align}
\tilde\mu(\bar p) &= \lambda \bar p - \frac{\eta}{2}\,(\lambda^2+3)^{n}\,\phi^{4n-2}\,|\bar p|^{2n},
\label{eq:Lmrho_mutilde_param}\\
\tilde p(\bar p) &= \bar p + \frac{\eta}{2\lambda}\,(\lambda^2+3)^{n}\Big[\lambda-2n(\lambda+1)\Big]\,\phi^{4n-2}\,|\bar p|^{2n}.
\label{eq:Lmrho_ptilde_param}
\end{align}
Again, the variable $\phi$ does not allow for a local closure relation between $\tilde\mu$ and $\tilde p$ in general. In order that the ratio $\tilde\mu/\tilde p$ be a constant, the following relation must be satisfied
\begin{equation}
    (1-2n)(1+\lambda)=0,
\end{equation}
which also implies uniquely that $\tilde\mu=\lambda\tilde p$. Therefore, only the cosmological-constant-type solution and the case $n=\frac12$ satisfy this requirement. For \(n=1/2\), any nonzero linear physical pressure yields the same normalized closure \(\tilde\mu=\lambda\tilde p\), and the corresponding phase portrait is
therefore the GR one, although the relation between physical and effective variables remains piecewise continuous. The dust limit must be treated separately. In that case \(\tilde p\equiv0\), so the pressure variable no longer carries information about \(\tilde\mu\), and the planar closure in the \((\tilde K,\tilde p)\)
chart degenerates. Consistency of Eq.~\eqref{eq:Kp_p} with
\(\tilde p\equiv0\) then leaves only the vacuum solution. Thus, a static self-gravitating dust ``star'' cannot be supported without pressure, akin to GR.

The scenario changes with respect to the case above when $n=1$. For the physical dust case in the \(L_m=-\rho\) branch, Eqs.~\eqref{eq:rho_eff_general_Lag_rho}-\eqref{eq:p_eff_general_Lag_rho} lead to
\begin{equation}
\rho_{\rm eff}=\rho-\frac{\eta}{2}\rho^2,
\qquad
p_{\rm eff}=-\frac{\eta}{2}\rho^2.    
\end{equation}
Again, EMSG endows pressureless matter with a nonvanishing effective pressure whose sign is opposite to that of $\eta$ in the $L_m=p$ case [see Eq.\ \eqref{eq:rho_p_emsg_Lp_dust}]. Therefore, the qualitative analysis will be the same under the identification $\eta\rightarrow-\eta$, with $\eta<0$ being the physically reasonable branch now. 


\section{Metric interpretation, physical variables, and surface matching}
\label{sec:recovering_results}

The covariant and normalized systems derived above are useful for the qualitative organization of the stellar equations. However, physical stellar models are usually described in terms of metric functions, a mass profile, and the radial dependence of the physical density and pressure. The purpose of this section is therefore threefold. First, we show how the covariant EMSG system maps into the standard metric representation. Second, we clarify how the fixed points of the normalized phase space should be interpreted in metric variables. Third, we rewrite the stellar equations in terms of the physical fluid variables and discuss the appropriate surface condition.

The important point is that the fixed points of the normalized system should not be confused with stellar surfaces at finite radius. A regular stellar configuration is an orbit in the covariant phase space. It starts from a central asymptotic regime and ends at a matching surface. In EMSG, this surface is determined naturally by the vanishing of the effective radial
pressure, \(p_{\rm eff}(R)=0\), rather than necessarily by \(p(R)=0\).

\subsection{Metric representation and effective TOV equation}
\label{sec:metric_TOV_map}

For a static, spherically symmetric geometry, the \(1+1+2\) variables are related to the usual metric functions by the line element
\begin{equation}
ds^2 = -e^{2\Phi(r)}dt^2 + \left(1-\frac{2m_{\rm eff}(r)}{r}\right)^{-1}dr^2 + r^2\left(d\theta^2+f^2(\theta)d\varphi^2\right),
\label{eq:metric_effective_star}
\end{equation}
where $m_{\rm eff}$ is the astrophysical mass, \(f(\theta)=\{\sin\theta,1,\sinh\theta\}\) for spherical, planar, or hyperbolic two-sheets, respectively. The corresponding covariant variables are \cite{Naidu:2021,Campbell:2025}
\begin{equation}
\phi = \frac{2}{r}\sqrt{1-\frac{2m_{\rm eff}}{r}},
\qquad
\mathcal{A} = \frac{d\Phi}{dr} \sqrt{1-\frac{2m_{\rm eff}}{r}},
\qquad 
K = -\frac{1}{r^2}\frac{1}{f}\frac{d^2 f}{d\theta^2}.
\label{eq:metric_covariant_dictionary}
\end{equation}
In the spherical stellar case, \(f(\theta)=\sin\theta\) and \(K=1/r^2\).

The effective mass function is defined by
\begin{equation}
m_{\rm eff}(r) = \frac{1}{2} \int_0^r \rho_{\rm eff}(\bar r)\,\bar r^{\,2}\,d\bar r,
\label{eq:meff_def}
\end{equation}
so that
\begin{equation}
\frac{dm_{\rm eff}}{dr} = \frac{1}{2}r^2\rho_{\rm eff}.
\label{eq:meff_derivative}
\end{equation}
Moreover, the hat derivative and the coordinate radial derivative are related by
\begin{equation}
\hat X = \sqrt{1-\frac{2m_{\rm eff}}{r}}\, \frac{dX}{dr}.
\label{eq:hat_radial_derivative}
\end{equation}
Using Eq.~\eqref{eq:pressure_hat_EMSG}, one obtains
\begin{equation}
\frac{dp_{\rm eff}}{dr} = - \frac{(\rho_{\rm eff}+p_{\rm eff})
\left[2m_{\rm eff}(r)+r^3p_{\rm eff}(r)\right]}{2r\left[r-2m_{\rm eff}(r)\right]}.
\label{eq:TOV_eff_coordinate}
\end{equation}
This is the usual TOV equation, but written for the effective density and pressure. In the GR limit, \(\eta\to0\), or in the weak-coupling regime \(|\eta\mathcal{T}^{\,n-1}|\ll1\), one has
\(\rho_{\rm eff}\to\rho\) and \(p_{\rm eff}\to p\), and
Eq.~\eqref{eq:TOV_eff_coordinate} reduces to the ordinary GR hydrostatic equilibrium equation. For the quadratic EMSG model, \(n=1\), it reproduces in covariant form the modified TOV equation previously obtained in the metric approach \cite{Nari:2018,Akarsu:2018b}.

The normalized effective variables are
\begin{equation}
\tilde K = \frac{1}{4} \left(1-\frac{2m_{\rm eff}}{r}\right)^{-1},
\qquad
\tilde p = \frac{p_{\rm eff}r^2}{4} \left(1-\frac{2m_{\rm eff}}{r}\right)^{-1},
\qquad
\tilde\mu = \frac{\rho_{\rm eff}r^2}{4} \left(1-\frac{2m_{\rm eff}}{r}\right)^{-1}.
\label{eq:metric_norm_cov_dictionary}
\end{equation}
Therefore, whenever \(\tilde K\neq0\), the inverse map gives
\begin{equation}
\mathcal{C}_{\rm eff} \equiv \frac{2m_{\rm eff}}{r} =
1-\frac{1}{4\tilde K},
\qquad
p_{\rm eff} = \frac{\tilde p}{\tilde K\,r^2}, 
\qquad
\rho_{\rm eff} = \frac{\tilde\mu}{\tilde K\,r^2}.
\label{eq:metric_inverse_map}
\end{equation}
Thus \(\tilde K\) is an algebraic measure of the effective compactness \(\mathcal C_{\rm eff}\), whereas \(\tilde p\) and \(\tilde\mu\) encode the effective pressure and density after normalization by the sheet expansion.

The map \eqref{eq:metric_inverse_map} clarifies how the finite fixed points of the normalized system should be interpreted. Since the normalized variables divide the physical source variables by powers of \(\phi\), a fixed point does not generally correspond to a stellar surface at finite radius. Instead, fixed
points describe asymptotic regimes of the normalized flow. Stellar
configurations correspond to trajectories in phase space, whose endpoint is fixed by a boundary condition in the metric variables.

For the linear effective closure $\tilde\mu=\lambda\tilde p$, the finite fixed points of the GR benchmark, or of any EMSG sector equivalent to it after the replacement \(\lambda\to\lambda_{\rm eff}\), can be translated into metric quantities as shown in Table~\ref{tab:metric_norm_cov_dict}. The table should not be read as a list of stellar surfaces. Rather, it indicates which metric regimes are associated with the fixed points of the normalized flow.

\begin{table}[ht]
\centering
\begin{tabular}{|c|c|c|c|}
\hline
Fixed point
&
$\mathcal{C}_{\textrm{eff}}$
&
$p_{\textrm{eff}}$
&
Metric interpretation
\\
\hline\hline
\(P_0\)
&
undefined
&
undefined
&
outside the regular spherical sector
\\
\hline
\(P_{1/4}\)
&
\(0\)
&
finite
&
regular-center or asymptotic vacuum
\\
\hline
\(\displaystyle P_{\rm int}\)
&
\(\displaystyle
\frac{4\lambda}{\lambda^2+6\lambda+1}
\)
&
\(\displaystyle
\frac{4}{(\lambda^2+6\lambda+1)r^2}
\)
&
self-similar singular scaling regime
\\
\hline
\(\displaystyle P_{K0} \)
&
undefined
&
undefined
&
non-spherical sector (not a star)
\\
\hline
\end{tabular}
\caption{Metric interpretation of the finite fixed points of the normalized linear-closure system. The table does not identify stellar surfaces. Fixed points describe asymptotic regimes of the normalized flow, while a stellar surface is determined by the matching condition \(p_{\rm eff}(R)=0\). The entries with \(\tilde K=0\) are singular in the inverse spherical metric map \eqref{eq:metric_inverse_map} and therefore do not represent regular finite-radius spherical stellar states.}
\label{tab:metric_norm_cov_dict}
\end{table}

In particular, \(P_{1/4}\) is the natural normalized value approached by a
regular center with finite central density and pressure, since
\(m_{\rm eff}(r)=O(r^3)\), \(\mathcal{C}_{\rm eff}\to0\),
\(\tilde K\to1/4\), and \(\tilde p\to0\) as \(r\to0\). By contrast, a stellar
surface at finite radius satisfies \(p_{\rm eff}(R)=0\), but it is not
generically a fixed point of the normalized autonomous system. At the surface,
one has
\begin{equation}
\tilde p(R)=0,
\qquad
\tilde K(R)
=
\frac{1}{4}
\left(1-\frac{2M}{R}\right)^{-1},
\label{eq:surface_normalized_values}
\end{equation}
with \(M=m_{\rm eff}(R)\). Thus, except in the zero-mass limit, the stellar
surface does not coincide with \(P_{1/4}\). This distinction is important in
EMSG, because the surface condition is imposed on the effective pressure.

\subsection{Stellar structure in terms of physical variables}
\label{sec:stellar_structure}

Although the covariant TOV equations are most naturally written in terms of
the effective quantities, stellar models are usually parametrized by the
physical density \(\rho(r)\) and pressure \(p(r)\). For this reason it is useful
to rewrite the structure equations directly in terms of the physical fluid,
keeping the EMSG corrections encoded algebraically in
\(\rho_{\rm eff}(\rho,p)\) and \(p_{\rm eff}(\rho,p)\).

The basic system consists of Eq.~\eqref{eq:TOV_eff_coordinate} together with
\begin{equation}
\frac{dm_{\rm eff}}{dr}
=
\frac{1}{2}r^2\rho_{\rm eff}(\rho,p),
\label{eq:mass_physical_system}
\end{equation}
supplemented by the physical EoS \(p=p(\rho)\). Inserting the explicit
expressions for \(\rho_{\rm eff}\) and \(p_{\rm eff}\), one obtains
\begin{equation}
\frac{dp}{dr}
=
-
\frac{
\left[\rho_{\rm eff}(\rho,p)+p_{\rm eff}(\rho,p)\right]
\left[2m_{\rm eff}(r)+r^3p_{\rm eff}(\rho,p)\right]
}
{
2r\,[r-2m_{\rm eff}(r)]
}
\left(
\frac{dp_{\rm eff}}{dp}
\right)^{-1},
\label{eq:dpdr_physical}
\end{equation}
where the derivative is taken along the physical EoS branch,
\begin{equation}
\frac{dp_{\rm eff}}{dp} = 1 + \frac{\partial p_{\rm emsg}}{\partial\rho} \frac{d\rho}{dp} + \frac{\partial p_{\rm emsg}}{\partial p} + \frac{\partial p_{\rm emsg}}{\partial c_s^2} \frac{dc_s^2}{dp}.
\label{eq:dpeff_dp}
\end{equation}
Equivalently, one may use the EoS to rewrite the system in terms of
\(\rho(r)\). The form \eqref{eq:dpdr_physical} assumes that
\(dp_{\rm eff}/dp\neq0\) along the branch under consideration. If this
derivative vanishes, the physical pressure ceases to be a good radial variable
for the effective TOV equation, and one should instead evolve a different
variable, such as \(\rho\) or \(p_{\rm eff}\) itself.

The central boundary conditions retain the standard form
\begin{equation}
m_{\rm eff}(0)=0,
\qquad
\rho(0)=\rho_c,
\qquad
p(0)=p(\rho_c),
\label{eq:central_data}
\end{equation}
and the equations are integrated outward. In this formulation the effect of
EMSG is entirely encoded in the algebraic map
\begin{equation}
(\rho,p)\longmapsto(\rho_{\rm eff},p_{\rm eff}),
\label{eq:effective_map_stellar}
\end{equation}
while the geometrical structure of the stellar problem remains GR-like. This
is one of the main advantages of the effective-fluid description: the same
stellar-integration strategy used in GR can be retained, but with source terms
modified by \(\eta\), \(n\), the physical EoS, and the choice of matter
Lagrangian.

\subsection{Matching conditions and radius definition}
\label{sec:matching_radius}

A particularly simple feature of EMSG is that it coincides with GR in vacuum. When \(T_{\mu\nu}=0\), one has \(\mathcal{T}=0\), so the matter-squared corrections vanish. The exterior geometry is therefore Schwarzschild,
\begin{equation}
ds^2
=
-\left(1-\frac{2M}{r}\right)dt^2
+
\left(1-\frac{2M}{r}\right)^{-1}dr^2
+
r^2d\Omega^2,
\label{eq:Schwarzschild_exterior_EMSG}
\end{equation}
with $M=m_{\rm eff}(R)$ being the total effective mass inside the star.

In the absence of surface layers, matching the interior solution to the Schwarzschild exterior is governed by the usual Israel-Darmois conditions. In the effective-fluid formulation, the natural radial-pressure condition at the surface is
\begin{equation}
p_{\rm eff}(R)=0.
\label{eq:surface_condition_peff}
\end{equation}
This condition is the effective-fluid counterpart of the usual GR condition \(p(R)=0\). For ordinary barotropic EoSes for which the vanishing of pressure at the surface also implies that the energy density goes to zero, the EMSG corrections vanish at the boundary, and the two definitions of the radius coincide. However, for self-bound matter, one may
have \(\rho(R)\neq0\) when the physical pressure vanishes. In that case, \(p(R)=0\) and \(p_{\rm eff}(R)=0\) are generally inequivalent.

It is therefore useful to distinguish two possible radii:
\begin{equation}
R_{\rm phys}:\quad p(R_{\rm phys})=0,
\qquad
R_{\rm eff}:\quad p_{\rm eff}(R_{\rm eff})=0.
\label{eq:two_radius_definitions}
\end{equation}
The radius relevant for smooth matching to the Schwarzschild exterior in the effective-fluid description is \(R_{\rm eff}\).  If \(R_{\rm phys}=R_{\rm eff}\), the usual GR intuition is recovered. If not, the stellar surface inferred  from the physical pressure differs from the surface selected by the gravitational source. In this case, \(R_{\rm eff}\) should be understood as the gravitational matching radius, whereas the
material interpretation of the region between \(R_{\rm phys}\) and
\(R_{\rm eff}\) depends on the chosen matter model. For instance, it may require a self-bound equation of state, an explicit cutoff prescription, or the inclusion of a surface layer if one insists on matching at the physical pressure surface.  This distinction is specific to the effective-fluid interpretation of EMSG and should be kept explicit in numerical integrations, mass--radius relations, and comparisons with previous metric treatments of
compact stars in EMSG.


\section{Conclusions and outlook}
\label{sec:conclusions}

In this paper we have constructed a covariant formulation of static, spherically symmetric stellar models in EMSG using the \(1+1+2\) semi-tetrad formalism. For the class of theories defined by \(F(R,\mathcal{T})=R+\eta \mathcal{T}^n\), we showed that the EMSG corrections generated by a perfect physical fluid can be reinterpreted as an effective perfect fluid. As a consequence, the stellar field equations take the GR form, but with \(\rho\rightarrow \rho_{\rm eff}(\rho,p)\) and \(p\rightarrow p_{\rm eff}(\rho,p)\).

This effective-fluid viewpoint yields a covariant TOV system and a dimensionless dynamical-systems formulation that are all formally identical to their GR counterparts once written in terms of the effective variables. The qualitative analysis shows that the reduced planar system is highly informative whenever a closure relation \(\tilde\mu=\tilde\mu(\tilde p)\) exists. In those cases, nullclines, invariant sets, finite equilibria, and fixed points at infinity provide a global organization of the stellar phase space. This makes it possible to identify the sectors in which EMSG is qualitatively or not equivalent to GR, those in which it reduces only sectorwise to a GR benchmark, and those in which the reduction to a planar system fails and the full three-dimensional covariant flow must be used instead. The dust sectors illustrate this last possibility particularly clearly.

From the point of view of previous results, the present approach recovers the standard GR stellar equations in the limit \(\eta\to0\), extends the known modified TOV equation beyond the quadratic EMSG model, and connects naturally with the GR dynamical-systems literature on relativistic stars. It also clarifies some special exact recoveries, such as the radiation sector for \(\mathcal{L}_m=p\), \(n=1\), and the \(n=\tfrac12\) sectors, where the effective closure becomes exactly or piecewise linear. The metric reconstruction further clarifies how the normalized fixed points should be interpreted: they organize asymptotic regimes of the effective flow, whereas stellar surfaces are selected by the matching condition.

A second important conclusion concerns the stellar surface. Because EMSG coincides with GR in vacuum, the exterior solution remains Schwarzschild. However, the natural matching condition is \(p_{\rm eff}(R_{\rm eff})=0\), not necessarily \(p(R_{\rm eff})=0\) or \(p(R_{\rm phys})=0\). For ordinary equations of state, the distinction may be negligible, but for self-bound matter it is genuine and should be kept explicit. This issue is not merely technical: it is one of the places where the effective-fluid description has direct physical consequences for the construction and interpretation of stellar models. A detailed perturbative and radial-stability analysis would be useful to
assess the observational consequences of this distinction.

The next step is to confront the present formalism with realistic EoSes and observable stellar quantities. In particular, it will be important to compute mass-radius relations, maximum masses, compactness bounds, and tidal deformabilities for representative neutron-star and quark-star equations of state, both in the quadratic model and in the generalized \(n\neq1\) case. Likewise, the covariant framework developed here provides a natural starting point for a radial-stability analysis, where the distinction between \(p(R)=0\) and \(p_{\rm eff}(R)=0\) may again become relevant.


\section*{Acknowledgements}
EB is partially supported by CNPq (grant N.\ 305217/2022-4) and FAPEMIG (APQ-05207-23). PKSD is supported by a grant from First Rand Bank (SA). SEJ is partially supported by FAPERJ.


\section*{References}
\nocite{*}

\bibliography{ref.bib}
\end{document}